\documentclass[%
preprintnumbers,
  amsmath,amssymb,
  aps,
]{revtex4-2}

\usepackage[utf8]{inputenc}

\usepackage{graphicx}
\usepackage{dcolumn}
\usepackage{bm}
\usepackage[dvipsnames]{xcolor}

\usepackage[normalem]{ulem}

\usepackage[version=4]{mhchem}

\usepackage{siunitx}
\DeclareSIUnit\bar{bar}

\usepackage{physics}

\usepackage[hidelinks]{hyperref}

\begin{document}

\title{Exploring extreme thermodynamics in nanoliter volumes through stimulated Brillouin-Mandelstam scattering}

\author{Andreas Geilen $^{1,2,}$\footnote[1]{These authors contributed equally.}}
\author{Alexandra Popp$^{1,2,3,\rm{a}}$}
\author{Debayan Das$^{1,4}$} 
\author{Saher Junaid$^{5,6}$}
\author{Christopher G. Poulton$^{7}$} 
\author{Mario Chemnitz$^{6,8}$}
\author{Christoph Marquardt$^{1,2,3}$}
\author{Markus A. Schmidt$^{5,6}$}
\author{Birgit Stiller$^{1,2}$}
\email{birgit.stiller@mpl.mpg.de}

\address{
$^{1}$Max Planck Institute for the Science of Light, Staudtstr. 2, 91058 Erlangen, Germany \\
$^{2}$Department of Physics, Friedrich-Alexander Universit\"at Erlangen-N\"urnberg, Staudtstr. 7, 91058 Erlangen, Germany \\
$^{3}$SAOT, Graduate School in Advanced Optical Technologies, Paul-Gordan-Str. 6, 91052 Erlangen, Germany\\
$^{4}$Universit\'e Bourgogne France-Comt\'e, 25030 Besan\c{c}on, France\\
$^{5}$Leibniz Institute of Photonic Technology, Albert-Einstein-Str. 9, 07745 Jena, Germany\\
$^{6}$Otto Schott Institute of Materials Research (OSIM), Fraunhoferstr. 6, 07743 Jena, Germany \\
$^{7}$School of Mathematical and Physical Sciences, University of Technology Sydney, NSW 2007, Australia\\
$^{8}$ INRS-EMT, 1650 Boulevard Lionel-Boulet, Varennes, Québec, J3X 1S2, Canada}

\date{\today}

\begin{abstract}
\noindent
Examining the physical properties of materials - particularly of toxic liquids - under a wide range of thermodynamic states is a challenging problem due to the extreme conditions the material has to be exposed to. Such temperature and pressure regimes, which result in a change of refractive index and sound velocity can be accessed by optoacoustic interactions such as Brillouin-Mandelstam scattering. Here we experimentally demonstrate Brillouin-Mandelstam measurements of nanoliter volumes of liquids in extreme thermodynamic regimes. We use a fully-sealed liquid-core optical fiber containing carbon disulfide; within this waveguide, which exhibits tight optoacoustic confinement and a high Brillouin gain of \SI{32.2\pm0.8}{\per\watt\per\meter}, we are able to conduct spatially resolved measurements of the Brillouin frequency shift. Knowledge of the local Brillouin response enables us to control the temperature and pressure independently over a wide range. We observe and measure the material properties of the liquid core at very large positive pressures (above \SI{1000}{\bar}), substantial negative pressures (below \SI{-300}{\bar}) and we explore the isobaric and isochoric regimes. 
The extensive thermodynamic control allows the tunability of the Brillouin frequency shift of more than \SI{40}{\percent} using only minute volumes of liquid. This work opens the way for future studies of liquids under a variety of conventionally hard-to-reach conditions.

\end{abstract}

\maketitle

\section{\label{sec:Intro}Introduction}

Understanding the physics at extreme thermodynamic points such as large pressure or high temperature has been a challenge for several decades \cite{Fortov.1997.Pureandappliedchemistry, Yoon.2020.PCCP}. The experimental study of liquids under such conditions can be especially demanding, due to the need for thermodynamic phase stabilization \cite{Holten.2017.TJOPCLett} and the handling of the liquid \cite{Carey.1998.JChemPhys}. Extreme thermodynamic experiments, such as those at high pressures and temperatures, become more and more demanding for safety reasons, particularly when toxic liquids are involved. Non-contact measurements are therefore highly desirable and can be performed by measuring optical \cite{Bhattacharya.1987.OpticsandLaserTechnology,Schunk.2014.OptExpress,Pumpe.2017.OptExpress} or acoustic \cite{Antman.2016.Optica, Marczak.1997.JASA, Zha.1994.PRB} parameters of the material. As an optoacoustic interaction, stimulated Brillouin-Mandelstam scattering (SBS) combines the advantages of both optics and acoustics and probes changes in temperature, pressure and density with high precision. SBS has been demonstrated in liquids \cite{Brewer.1964.PRL}, gases \cite{Hagenlocker.1965.APL} and in optical fibers \cite{Kobyakov.2010.Adv.Opt.Photon.}; fibers in particular are a very common platform for SBS investigations such as microwave photonics \cite{Zarifi.2021.OptLett}, signal processing \cite{Santagiustina.2013.SciRep} and sensing \cite{Zadok.2012.LaserPhotonicsReviews} because of their low optical loss, long interaction lengths, easy handling and tight optoacoustic confinement.  \\

Conventional optical fibers mainly consist of fused silica or highly nonlinear solid cores, which makes it almost impossible to study thermodynamic effects. Liquid-core optical fibers (LiCOF) open up access to thermodynamics of fluids while preserving the advantages of optical fibers, relying only on nanoliter volumes for meter long interaction lengths. This platform has already found applications in supercontinuum generation \cite{Chemnitz.2018.Optica}, lasing \cite{Kieu.2013.Opt.Lett.} temperature sensing \cite{Smith.1974.TheJournaloftheAcousticalSocietyofAmerica} or signal processing \cite{Chemnitz.2021.NovelOpt.Mat.Appl.}. Fully-sealed fibers elevate this idea to the next level by removing additional layers of system complexity while simultaneously opening up additional thermodynamic regimes. Therefore, fully-sealed LiCOFs together with SBS appear as an ideal tool to unlock new physical effects.
\\
\\
In this work, we experimentally study the interaction of optical and acoustic waves in fully-sealed LiCOF filled with only a few nanoliters of highly toxic carbon disulfide (\ce{CS2}). We choose \ce{CS2} as the core medium, since it is an especially promising platform due to its high nonlinear gain. Our fully-sealed, all-fiber setup allows for full control of the state of the volatile solvent \ce{CS2}. We thoroughly characterize the optoacoustic response of our fiber and support our measurements of the acoustic resonances with numerical simulations of the longitudinal acoustic modes present in the core. We identify a large Brillouin gain of \SI{32.2\pm0.8}{\per\watt\per\meter} with a Brillouin linewidth of \SI{65}{\mega\hertz}. To deepen our understanding, we perform spatially resolved measurements of the optoacoustic interaction using Brillouin optical correlation domain analysis (BOCDA) with cm-resolution. This investigation at the length scale of the thermodynamic processes allows us to determine the local and global influences of temperature and pressure changes.
\\
\\
With a single device, we probe up to pressure values larger than \SI{1000}{\bar} and down to negative pressure values smaller than \SI{-300}{\bar}. With these insights, we are able to perform thermodynamic control of the SBS, allowing us to tailor the Brillouin frequency shift (BFS) by more than \SI{40}{\percent} of its initial value with high precision. We extend our measurements from the isochoric (constant volume) to the isobaric (constant pressure) regime of the liquid. Studying the latter, we show that the absence of the pressure component leads to a more than 7-fold enhancement in sensitivity of the BFS with temperature changes compared to standard solid core fibers. In this case, SBS is extremely-well suited, since the air gap formation inside the fiber prevents any transmission measurements. Additionally, the isochoric state at the boundary to the isobaric state allows stable and reproducible access to the regime of negative pressure. This metastable state is associated with a stretched liquid, where attractive forces between the molecules prevent the formation of an energetically more favorable state of separated liquid and gas phase \cite{Imre.1998.JournalofNonEquilibriumThermodynamics}. This regime is normally only accessible with high experimental expenses and has rarely been achieved in liquids other than water and at such high negative pressure values \cite{CAUPIN20061000}. Indeed, probing those boundaries of metastable thermodynamic regimes is of particular importance for verifying the equations of states in common molecular models \cite{Zheng.1991.Science}. Our work uniquely showcases the use of optoacoustics as a tool to advance our understanding of thermodynamics as well as the potential to implement highly sensitive fiber sensors, optoacoustically monitored chemical nano-reactors and thermodynamically tunable radio-frequency filters.

\section{\label{sec:SBS_in_LiCOF} Brillouin-Mandelstam scattering in \ce{CS2}-filled LiCOF}
\begin{figure}[b]
\centering
\includegraphics[width=1\textwidth]{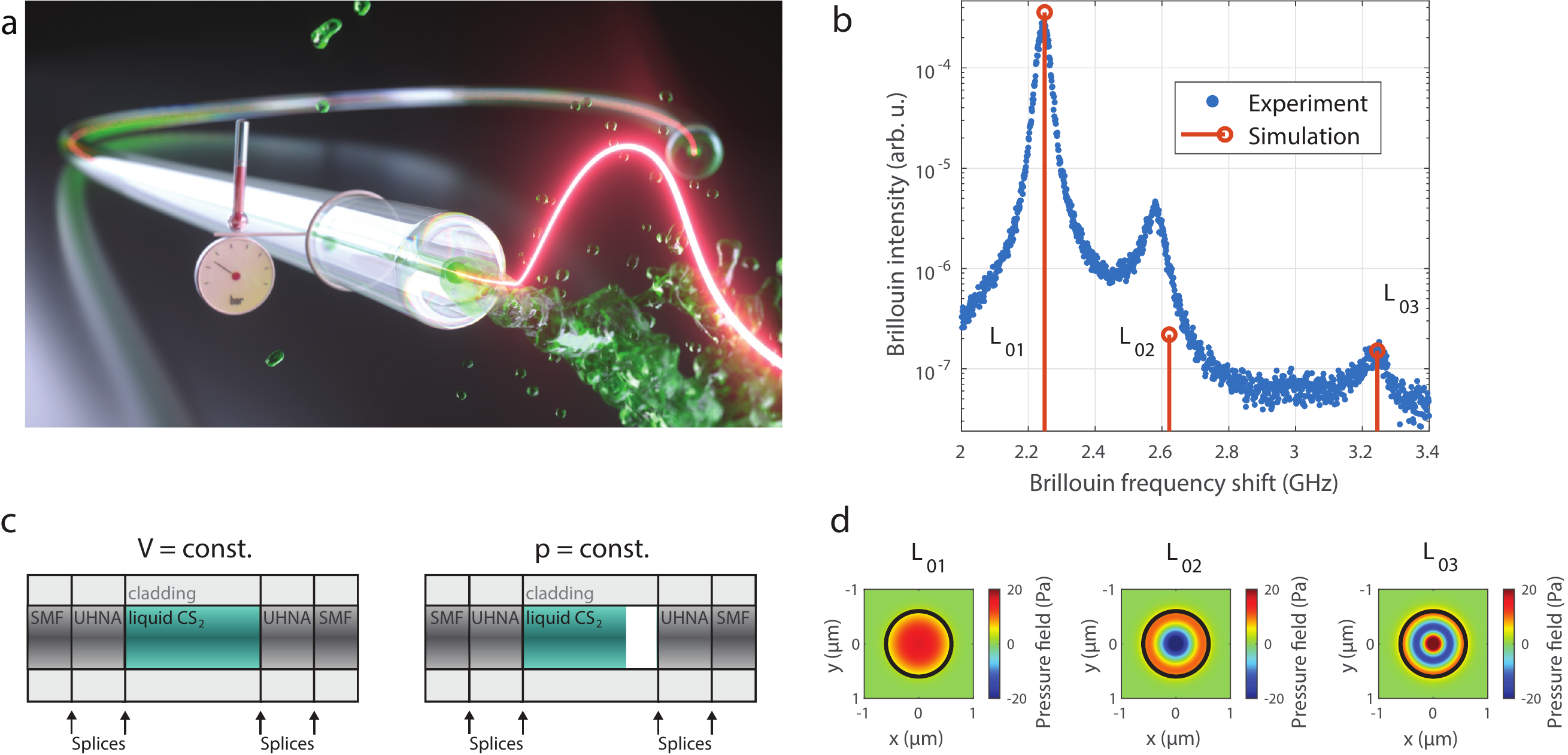}
\caption{ \textbf{a)} Artist's view of the LiCOF sample with laser. 
\textbf{b)} Measured spectrum (blue) and simulation (red) of stimulated Brillouin-Mandelstam scattering spectrum with \SI{1.125}{\micro\meter} core diameter.
\textbf{c)} Geometry of LiCOF sample. Spliced on both sides with ultra high numerical aperture (UHNA) fiber for coupling and sealing. SMF for integrability. If the capillary is fully-filled, the volume stays constant (isochoric regime) while if a bubble is present the pressure stays constant (isobaric regime).
\textbf{d)} Numerical simulations of corresponding longitudinal, radially symmetric pressure acoustic modes $\mathrm{L_{01}}$, $\mathrm{L_{02}}$ and $\mathrm{L_{03}}$, respectively.}
\label{fig:Introduction}
\end{figure}

Brillouin-Mandelstam scattering is an optoacoustic effect whereby a strong pump laser is coherently scattered by traveling acoustic waves, resulting in a back-scattered Stokes and anti-Stokes signal.
The Stokes signal is frequency downshifted by the Brillouin frequency shift $\Omega_\text{B}/2\pi$ and shows a Lorentzian shape. A comprehensive tutorial on Brillouin scattering is found in the work by Wolff~et~al.~\cite{Wolff.2021.J.Opt.Soc.Am.B}.
The Brillouin frequency shift is described by 
\begin{align}
\label{eq:delN}
\frac{\Omega_\mathrm{B}}{2\pi} & = \frac{2}{\lambda}  \cdot n_\mathrm{eff}\qty(p,T)  \cdot v_\mathrm{ac}\qty(p,T),
\end{align}
\noindent where both the speed of sound $v_\mathrm{ac}\qty(p,T)$ and the mode refractive index $n_\mathrm{eff}\qty(p,T)$ are dependent on the temperature and the pressure inside the fiber \cite{Hartog.2017.CRC}. 
In contrast to solid-core fibers, in LiCOFs externally applied local stresses distribute throughout the entire core in the form of changes in global pressure.
For the mode refractive index we use the extended Sellmeier equation from Chemnitz et~al.~\cite{Chemnitz.2021.NovelOpt.Mat.Appl.} to describe pressure and temperature dependence.
We assume that the speed of sound can be described by a linear combination of the contributions from the pressure and temperature separately, with the pressure contribution taken from the model of Smith~et~al.~\cite{Smith.1974.TheJournaloftheAcousticalSocietyofAmerica} and the temperature contribution from that of Eastman~et~al.~\cite{Eastman.1969.TheJournalofChemicalPhysics}.
In this case $v_{\mathrm{ac, CS2}}\qty(p= \SI{1}{\bar},T=\SI{20}{\celsius}) = \SI{1242}{\meter \per \second}$ is the ambient condition speed of sound of \ce{CS2}.
\\
\\
The liquid \ce{CS2} is sealed within a \ce{SiO2} glass capillary by spliced ultra high numerical aperture fiber (UHNA) and standard single mode fibers (SMF).
Different samples were used in the experiments which have slightly different parameters due to fabrication reasons.
Unless otherwise specified, in the following a length of $L = \SI{2\pm0.05}{\meter}$ and core diameter of \SI{1.125}{\micro\meter} are the parameters of the samples employed in this study.
Not only is the LiCOF optically guiding, but also acoustically: as the longitudinal velocity of sound in the core ($\approx$ \SI{1242}{\meter\per\second} \cite{Smith.1974.TheJournaloftheAcousticalSocietyofAmerica}) is less than the velocity of shear wave in the silica cladding ($\approx \SI{3765}{\meter\per\second}$), the acoustic mode fields in the cladding are all evanescent and the acoustic mode is therefore guided by total internal reflection.
More information on the samples is found in the methods section and the work of Chemnitz et~al.~\cite{Chemnitz.2017.NatCommun}.
\\
\\
Controlling the temperature in the capillary can be done straightforwardly by heating or cooling externally, but controlling the pressure directly is more involved. An increase in temperature leads to expansion of the \ce{CS2}, following $\left(\frac{\partial L}{\partial T}\right)_p = \alpha_\text{V}$, where $\alpha_\text{V} = \SI{1.12e-3}{\per \kelvin}$ is the volumetric coefficient of expansion of \ce{CS2} at \SI{20}{\celsius} \cite{Haynes.2016.}.
If the liquid fills the complete capillary, it cannot further expand, which then translates to an increase of pressure.
Following the work of Pumpe~et~al.~\cite{Pumpe.2017.OptExpress} and extending it for partially heated fibers, the relation between a temperature change and a pressure change is determined by 
\begin{equation}
    \label{eq:pressure_calc}
    \Delta p = \frac{\alpha_\text{V}}{\kappa} \frac{L_\mathrm{H}}{L}\Delta T,
\end{equation}
where $\kappa = \SI{9.38e-10}{\per\pascal}$ the isothermal compressibility of \ce{CS2} \cite{Haynes.2016.} and $L_\mathrm{H}$ is the heated length, assuming $\alpha_\mathrm{V} \Delta T \ll 1$.
In contrast to non-sealed systems \cite{Kalogerakis.2007.JOSAB, Behunin.2019.PRA}, this relation provides full control of the pressure without requiring open access to the liquid, by partially heating or cooling the fiber while keeping the temperature of the other region constant.
\\
\\
To illustrate the thermodynamic behavior of this system, the schematic geometry of the LiCOF is shown in \autoref{fig:Introduction}c.
Two regimes are depicted in \autoref{fig:Introduction}c:
First the isochoric regime, in which the fiber is fully-filled and the \ce{CS2} volume is constant (upper graphic). 
In this regime the fiber has optical transmission and an increase of temperature also increases the internal pressure.
Second, the isobaric regime, where a gap is present and the pressure in the \ce{CS2} is constant (lower graphic).
In this regime the fiber does not transmit light but is accessible via backward SBS from one side.

\section{\label{sec:Integrated} Integrated Brillouin-Mandelstam analysis}
\begin{figure}[t]
\centering
\includegraphics[width=1\textwidth]{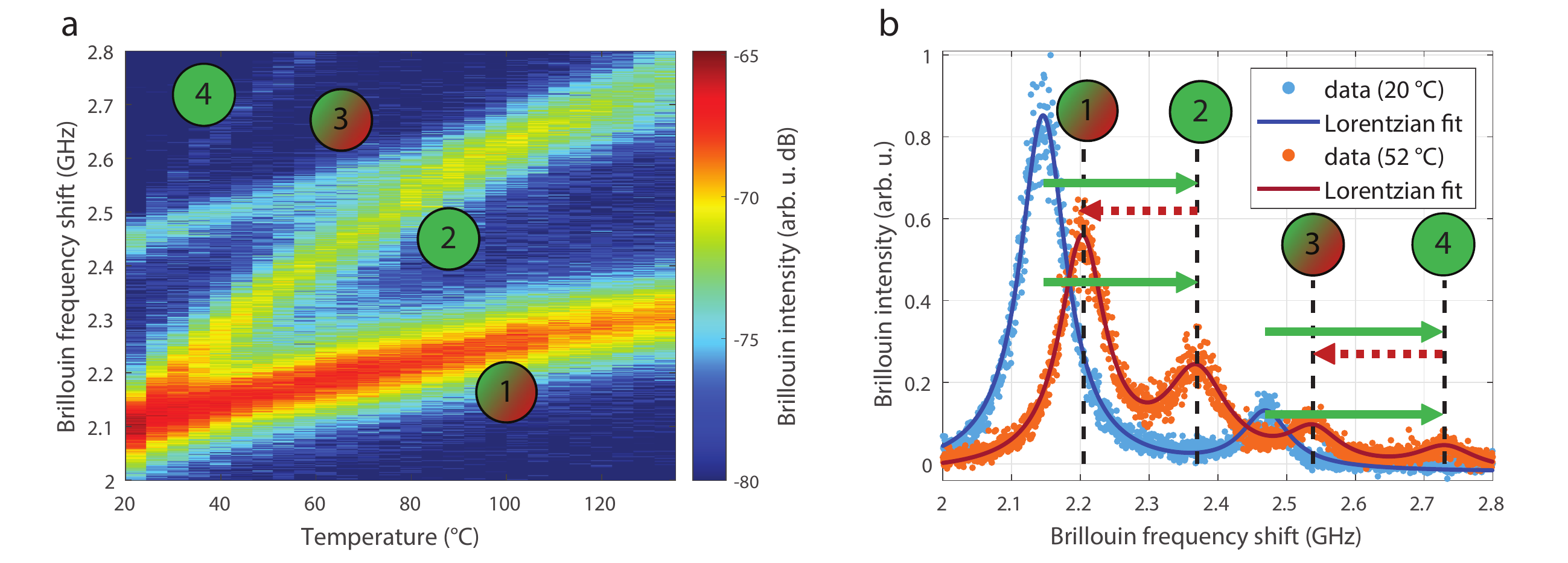} 
\caption{Explanation of the observed splitting in integrated Brillouin measurements.
\textbf{a)} Map of SBS response to partial heating. A splitting of the optical modes related to the first (second) acoustic mode into two modes 1 and 2 (3 and 4) is observed. \textbf{b)} Global influence pressure (up-shift - green) and local influence of temperature (down-shift - red) act inversely on the BFS. While pressure acts on all modes, only mode 1 and 3 are influenced by temperature directly. The spectra match a multi-Lorentzian fit.}
\label{fig:Integrated}
\end{figure}

In order to gain fundamental understanding of the Brillouin-Mandelstam response of the LiCOF, we firstly investigate it via an integrated, thermally seeded SBS analysis at a pump wavelength of $\lambda = \SI{1549}{\nano\meter}$.
The SBS spectrum at room temperature is shown in \autoref{fig:Introduction}b.
Numerical simulations reveal that the peaks are due to high optoacoustic overlap between the fundamental optical mode and the first three radially symmetric longitudinal pressure acoustic modes $\mathrm{L_{01}}$, $\mathrm{L_{02}}$ and $\mathrm{L_{03}}$ (\autoref{fig:Introduction}d).
Thus, the simulations confirm the origin of the rich mode spectrum due to multiple acoustic modes only, in the absence of higher order optical modes.
For the peak at \SI{2.25}{\giga \hertz} we find a total gain of \SI{9.29 \pm 0.23}{\per \watt \per \meter}, and a gain of \SI{32.2 \pm 0.8}{\per \watt \per \meter} when compensating for the coupling losses of \SI{5.4}{\decibel} with a linewidth of \SI{65}{\mega\hertz}.
This value is two orders of magnitude higher than the values of standard SMF (around 0.2 - \SI{0.3}{\per \watt \per \meter} \cite{Lanticq.2009.OptLett, Nikles.1997.JLWTech}).
\\
\\
In \hyperref[fig:Integrated]{Figure~\ref*{fig:Integrated}}a, the temperature response of the LiCOF is investigated by measuring the Brillouin spectrum while heating a part of the capillary from room temperature to \SI{130}{\celsius} in steps of \SI{5}{\celsius} on a hotplate. 
Above room temperature, the peaks associated with the first and second order longitudinal pressure acoustic modes, split up and move to higher frequencies with increasing temperature.
To identify the modes in the following, we number them from $1$ to $4$ by increasing frequency. 
The observed splitting is explained in \hyperref[fig:Integrated]{Figure~\ref*{fig:Integrated}}b by the additive but opposite influence of global pressure and local temperature. 
First, the global pressure influence up-shifts the BFS inside and along the whole core, which is depicted by the green solid arrows in \autoref{fig:Integrated}b.
Second, the local temperature influence down-shifts the BFS only in the heated part of the capillary, which is depicted by the red dashed arrows.
In the integrated spectrum, the local Brillouin spectra along the whole fiber are summed up \cite{Kobyakov.2010.Adv.Opt.Photon.}, resulting in the observed splitting in \autoref{fig:Integrated}a.
We can now identify each of the modes with the influences they are subjected to:
While all four modes are subject to pressure changes, only modes $1$ and $3$ are subject to temperature changes.
This allows us to discriminate the influences of both temperature and pressure.


\section{\label{sec:Distributed} Spatially resolved Brillouin-Mandelstam analysis}

%
\begin{figure}[t]
\centering
\includegraphics[width=1\textwidth]{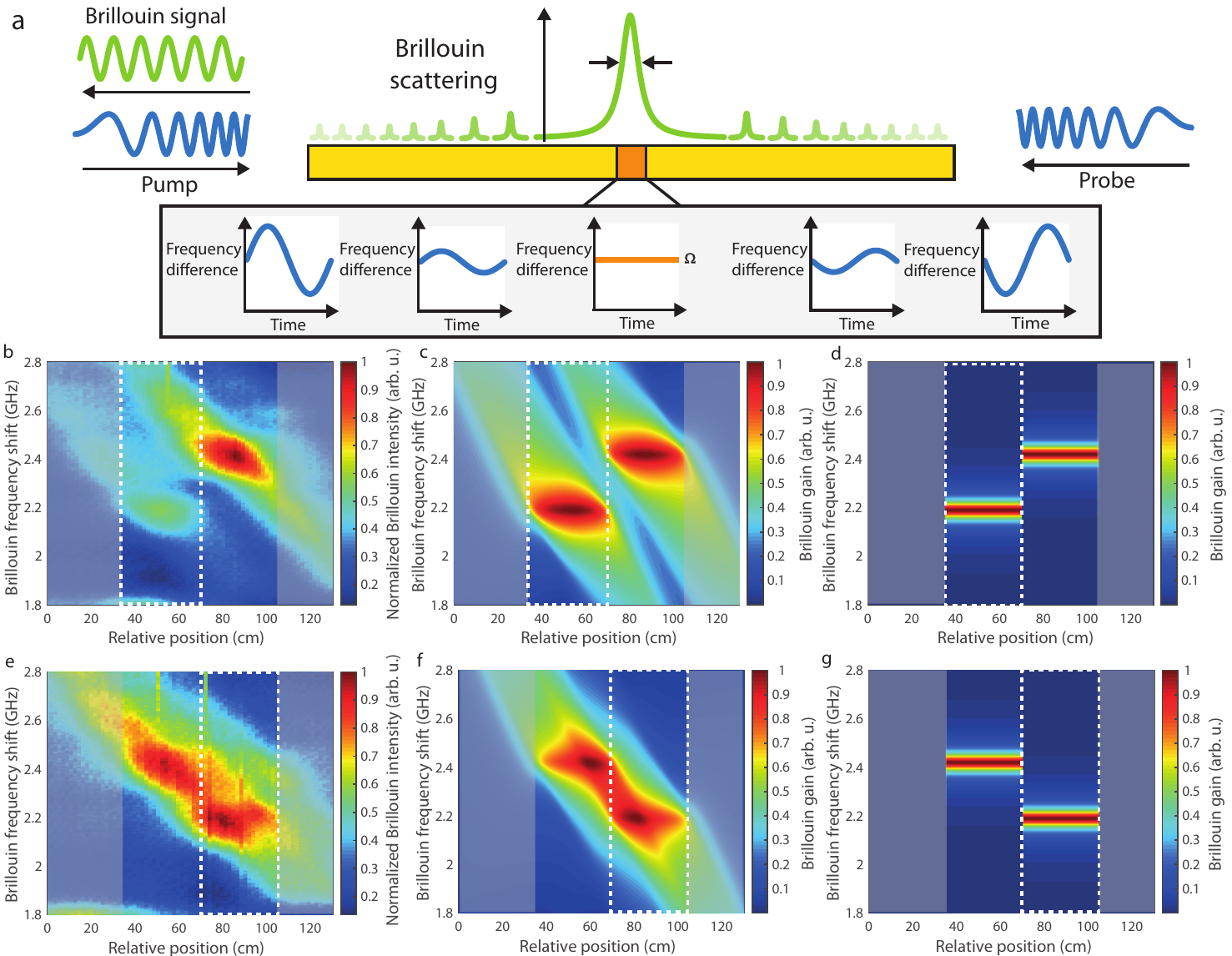}
\caption{Distributed Brillouin measurements \textbf{a)} Scheme of Brillouin optical correlation domain analysis (BOCDA). Frequency modulated, counter propagating pump and probe wave are launched into the fiber under test (FUT), causing a time and position dependent frequency beating. Only at specific correlation positions, the acoustic wave is stimulated emitted continuously, leading to a continuous-wave, localized Brillouin response. \textbf{b)} and \textbf{e)} Position resolved measurement of Brillouin frequency shift in partially heated LiCOF, Brillouin intensity color coded. Heated length is denoted by dashed white box. 
Shaded grey area denotes SMF system parts. Pump direction is from the right. BFS in the non-heated part is up-shifted compared to the heated part by approximately $\Delta \nu = 200\,$MHz. \textbf{c)} and \textbf{f)} Numerical simulations of heated LiCOF BFS response. \textbf{d)} and \textbf{g)} Initial Brillouin gain distribution retrieved from simulations.}
\label{fig:BOCDA}
\end{figure}
To experimentally confirm the local and global nature of the pressure and temperature influence on the different modes, we employ direct-frequency modulated Brillouin optical correlation domain analysis (BOCDA) \cite{Hotate.2000.IEICEtransactionsonelectronics} for a distributed analysis of the fiber. 
Following the scheme in \autoref{fig:BOCDA}a, a frequency modulated, counter propagating pump in combination with a probe creates a localized acoustic response inside the fiber. By changing the frequency of the modulation, the correlation position can be moved throughout the fiber to create a position dependent map of the BFS. The resolution and sensing range are governed by the modulation frequency $f_m = 699\,$kHz and the modulation bandwidth $\Delta f = \SI{47}{\giga\hertz}$, using 
\begin{align}
\Delta z = \frac{\Delta\nu_\text{B} \, v_\text{g}}{2\pi f_m \Delta f}.
\label{eq:BOCDA:resolution}
\end{align}

This yields a resolution $\Delta z = 6.4\,$cm and sensing distance $d_m = 146\,$m for a Brillouin gain bandwidth $\Delta\nu_B = \SI{65}{\mega\hertz}$, and a group velocity of $v_\text{g} = c/n$ with an assumed refractive index of $n = 1.47$ arising from the simulations of section \autoref{sec:Integrated}. Due to the longer measurement times of the distributed approach, shorter fibers of $L = \SI{0.8\pm0.05}{\meter}$ are used in this section.\\

Similar to previous measurements, we heat several fractions of the fiber up to \SI{130}{\celsius} and record the BFS response. \hyperref[fig:BOCDA]{Figure~\ref*{fig:BOCDA}}b and e show typical results, with either the half at the incoupling of the probe (\autoref{fig:BOCDA}b) or half at the incoupling of the pump (\autoref{fig:BOCDA}e) heated to \SI{80}{\celsius}. The shaded parts denote the SMF and UHNA. The SMF response at these positions is not shown, due to the BFS of $\nu_B = 10.8\,$GHz. The dashed white box denotes the region which is heated, pump direction is always the right side of the figure. Calibration measurements reveal a room temperature BFS of $\nu_B = 2.2\,$GHz, in agreement with the integrated data. In contrast to the integrated measurements, only the dominant SBS mode is visible, due to the reduced signal-to-noise ratio of the distributed approach. Furthermore, all measurements exhibit a strong asymmetric background, which is a common feature for direct-frequency modulated BOCDA \cite{Song.2018.OptLett} and arises from the non-instantaneous amplitude response of the frequency modulated laser source. To treat this we conduct numerical simulations for the Brillouin gain following references \cite{Song.2018.OptLett} and \cite{Yamauchi.2004.Fiber.Opt.Sens.Tech.} shown in \autoref{fig:BOCDA}c and f. Neglecting the pump-loss inside the fiber, the simulations clearly reproduce the experimental data. The simulations retrieve the initial, undisturbed Brillouin gain distribution (\autoref{fig:BOCDA} d and g). Additionally, in the simulations we also suppress the correlation beat-noise, which is independent of the laser source. We find, that the BFS of the heated regions is steady at $\nu_B = 2.2\,$GHz for all temperatures, while the BFS of the non-heated regions is up-shifted. This is in agreement with the findings from \autoref{sec:Integrated} and confirms our theory of local temperature and global pressure influence in the fiber.


\section{\label{sec:Thermodynamics} Extreme thermodynamic regimes}

\begin{figure}[t]
\centering
\includegraphics[width=1.0\textwidth]{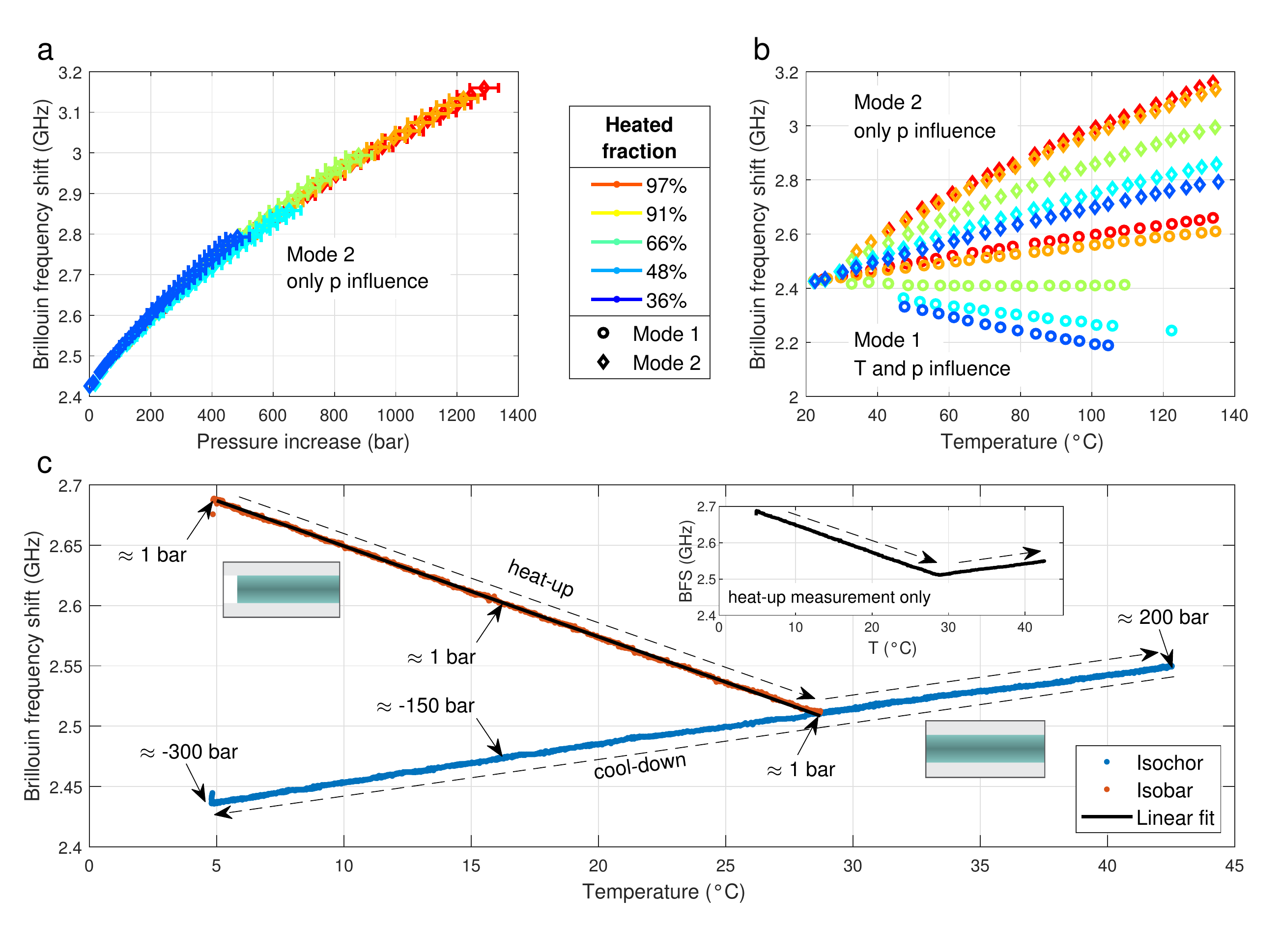}
\caption{Thermodynamic control of stimulated Brillouin scattering (Sample: \SI{2.1}{\micro\meter} diameter and \SI{2}{\meter} length)
\textbf{a)} BFS of mode $2$ over relative pressure increase for differently long heated fractions of the fiber.
\textbf{b)} Thermodynamic control of BFS of mode $1$ and mode $2$, which originate from the same resonance, by almost \SI{1}{\giga\hertz} (\SI{0.97}{\giga\hertz}), changing temperature and pressure inside the fiber core.
\textbf{c)} BFS of mode $1$ is observed in the isobaric (red curve) and isochoric (blue curve) regime by first cool-down of the LiCOF, then switching to the isobaric regime by cavitation due to local heating and finally heat-up of the LiCOF again above the recombination point.
In the isobaric regime, we find a large slope of \SI{-7.5}{\mega \hertz \per \celsius}.
In the isochoric regime negative pressures are achieved down to \SI{-300}{bar}.
In the inset, the heat-up data is shown, revealing the continuous switching from isobaric to isochoric regime around \SI{28}{\celsius}.
The anomaly around \SI{5}{\celsius} is due to the initiated cavitation. A p-T diagram is provided in the supplementary material.}
\label{fig:ThermodynamicControl}
\end{figure}

As stated in \autoref{eq:pressure_calc}, the pressure variation inside the fiber can be achieved using partial heating of the waveguide as an indirect method of control. In order to validate this assumption, the fiber is heated along different fractions and the BFS of mode $1$ and $2$ are measured. Mode 1 is dependant on temperature and pressure, while mode 2 is only dependent on the pressure inside the capillary. In \autoref{fig:ThermodynamicControl}a, the BFS of mode $2$ is plotted against the pressure change. The dominating uncertainty of the data points in \autoref{fig:ThermodynamicControl}a is caused by the uncertainty in the heated length.
The absolute value of the BFS differs from \autoref{fig:Integrated} because of different sample parameters (\SI{2.1}{\micro\meter} diameter instead of \SI{1.125}{\micro\meter} diameter and \SI{2}{\meter} length).
The overlap for the different heated fractions indicates that the pressure can be tuned independently from temperature by changing the length of the heated fraction of the capillary. 
We then complete this picture by including mode $1$, which is dependent on the pressure as well as the temperature as shown in \autoref{fig:ThermodynamicControl}b.
With the full understanding of the pressure and temperature dynamics of \autoref{eq:delN}, we are able to tailor the BFS of the fiber over a large range of almost \SI{1}{\giga\hertz} (\SI{0.97}{\giga\hertz}).
This is an unprecedented tunability of more than \SI{40}{\percent} of the initial value, using only fractional heating, which can be easily implemented with a standard hotplate.
\\
\\
In a next step, the fiber is cooled to temperatures below \SI{5}{\celsius} (\autoref{fig:ThermodynamicControl}c) resulting in a pressure reduction. 
In this measurement, most of the fiber is located on a Peltier element except for a short section that is attached on a hotplate. We therefore use mode $1$, which is influenced by both temperature and pressure.
Initially, the fiber is fully-filled and thus follows isochoric cooling, therefore, during cool-down, it shows the expected decrease of the BFS. When the inside of the capillary reaches vapor pressure of \ce{CS2} \cite{Waddington.1962.J.Phys.Chem.}, the BFS decrease continues with steady slope from the positive into the negative pressure regime.
The latter state may seem less intuitive as one might expect the pressure to be always positive. However, as previous works show, a liquid in a tight confinement, such as capillaries or micro-pores, can be stretched due to adhesion to the vessels boundaries, opening a new thermodynamic metastable regime \cite{Zheng.1991.Science,CAUPIN20061000}.
In this case, the liquid column is completely filled and the attractive forces between the molecules (cohesion) lead to an inner tension that is expressed as a negative pressure as soon as the system is cooled below the temperature where it would reach vapor pressure.
Under this condition, the energetically more favorable phase separation (isobaric state) is suppressed by the attractive cohesion force between the molecules \cite{Davitt.2010.TheJournalofChemicalPhysics}.
This metastable state of the liquid column remains intact until an inhomogeneity in the system or a fluctuation of the environment leads to nucleation (bubble formation). Here, as soon as a small perturbation is present, the liquid column will break up and a liquid-gas interface forms that transfers the system into an isobaric state. The gas phase then expands until vapour pressure and cohesive forces are in balance again.
The metastability of this particular thermodynamic regime is described by classical nucleation theory \cite{Debenedetti.1996.}, after which an energy barrier must be overcome to break the balance between the forces, eventually leading to homogeneous nucleation. In experiment, this statistical process is usually triggered by perturbations in the liquid or the experimental system \cite{CAUPIN20061000} which is called heterogeneous nucleation.
Additionally, it was shown by Brereton~et~al.~\cite{Brereton.1998.ChemicalPhysicsb}, that the smaller the capillary, the lower the pressure can get where heterogeneous nucleation occurs.
In our case, the small volume of \ce{CS2}, the high quality of the capillaries, and high mechanical strength of the splices allows to access negative pressures of less then \SI{-300}{\bar} following \autoref{eq:pressure_calc} without homogeneous nucleation.
\\
\\
In order to initiate nucleation in a controlled way, a small fraction at the end of the fiber is heated above the boiling point of \ce{CS2}.
At this point, the transmission of the sample is lost, confirming the breakup of the liquid column and thus the onset of the isobaric regime.
Furthermore, a strong jump in BFS is visible, caused by the sudden pressure change from large negative pressure to the vapor pressure of the \ce{CS2} inherent with nucleation. 
Next, the liquid is heated in the isobaric regime. The liquid expands with increasing temperature without relevant pressure influence on the BFS \cite{Waddington.1962.J.Phys.Chem.}.
This leads to a large negative slope of the BFS of \SI{-7.5}{\mega \hertz \per \celsius} (red curve of \autoref{fig:ThermodynamicControl}c).
The BFS decreases continuously with a linear behavior until it reaches the recombination point of \SI{28.7}{\celsius}. 
Here, the transmission of the sample is restored as the bubble is closing again. 
The fiber transitions back from the isobaric to the isochoric regime which are indistinguishable at the recombination point.
The exact temperature of the recombination point depends on the filling ratio of the fiber as well as ratio of heated to cooled length.
For temperatures above this recombination point, the behavior is as described in \autoref{sec:Integrated}.


\section{\label{sec:Conclusion} Conclusion}

In summary, we have investigated large positive and negative pressure regimes in the isobaric and isochoric liquid state of \ce{CS2} with help of integrated and spatially resolved Brillouin-Mandelstam scattering experiments in fully-sealed \ce{CS2}-filled capillary fibers and measured a large gain of \SI{32.2 \pm 0.8}{\per \watt \per \meter}. Extreme thermodynamic regimes of more than \SI{1000}{\bar} and less than \SI{-300}{\bar} were reached, which enables extraordinary control of the Brillouin response with a tunability of \SI{40}{\percent} using only a few nanoliters of the highly nonlinear but toxic and volatile liquid \ce{CS2}. Distributed Brillouin measurements with cm spatial resolution, which is at the length scale of the thermodynamic processes, allowed us to experimentally distinguish the local response of temperature changes and the global response of different pressure values. In the isobaric state, we were able to observe the purely temperature related BFS change of \SI{-7.5}{\mega \hertz \per \celsius}.
\\
\\
In contrast to previous studies \cite{Kieu.2013.Opt.Lett., Fanjoux.2017.JOSAB}, our work opens completely new optoacoustic insights into the thermodynamic properties of a liquid. The thermodynamic regimes that we studied can only with difficulty be attained in other platforms as the fully-sealed fibers allow to access extreme pressure values, different volumetric regimes of the liquid and independent temperature control in a reproducible manner. In addition, backward SBS provides unique features such a being able to measure in the isobaric regime where no optical transmission is possible and to provide a spatially resolved measurement of the refractive index and the sound velocity. 
\\
\\
Our results significantly contribute to the understanding of liquids in extreme thermodynamic states and give exclusive insights into the thermodynamics-related changes of the speed of sound and the refractive index in such. Furthermore, these results pave the way for a future generation of liquid-core Brillouin photonics, including but not limited to highly sensitive fiber-based monitoring devices, a platform for controlled nano-liter chemistry and widely tunable radio-frequency applications.

\section{Methods}

\subsection{LiCOF Sample}

The LiCOF sample is depicted in \autoref{fig:Introduction}c of the main text. It consists of a \SI{}{\micro \meter}-size silica capillary, which is filled with liquid \ce{CS2}.
Both ends are sealed by spliced ultra high numerical aperture (UHNA) fibers.
This provides efficient optical coupling and contains the \ce{CS2} inside the core.
Finally, on both sides standard SMF with FC/UPC connectors are spliced to allow simple implementation into the setup.
Optical and acoustic guidance is provided as the speed of sound as well as the speed of light in the \ce{CS2} core is lower as it is in the cladding. 
The core size of the samples used in this work is \SI{1.125}{\micro \meter} ($V = 1.51$, thus optically single mode at $\lambda = $\SI{1549}{\nano \meter}) and \SI{2.1}{\micro \meter} ($V = 2.82$, thus optically multi mode at $\lambda = $\SI{1549}{\nano \meter}). 
The length of the LiCOF part is between \SI{0.8}{\meter} and \SI{2.0}{\meter}.

\subsection{Setup for integrated SBS measurements \label{sec:LiCOF_int_setup}}

\begin{figure}[ht]
\centering
\includegraphics[width = 0.6\textwidth]{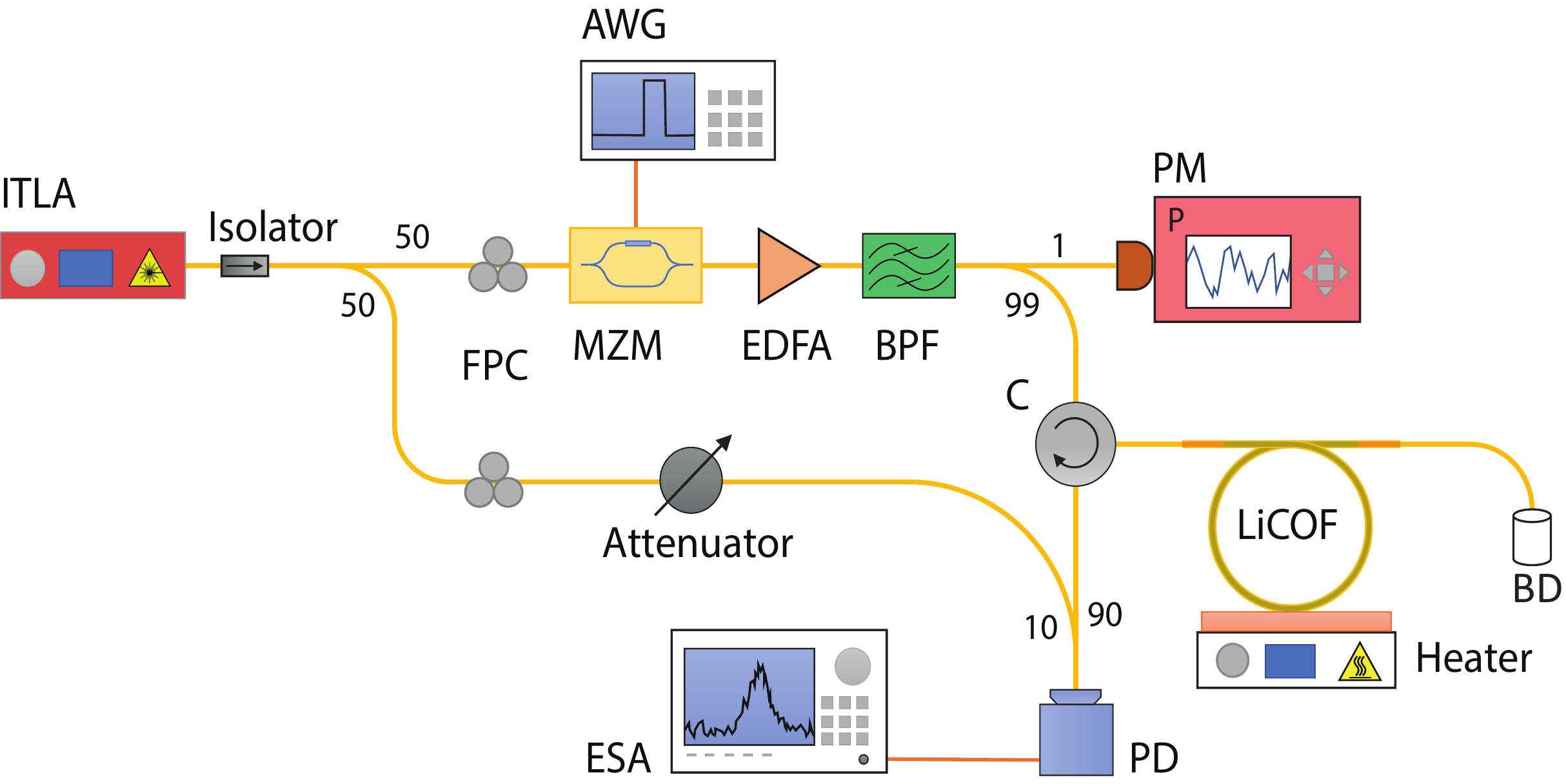}
\caption{Schematic sketch of integrated SBS setup. ITLA: integrated tunable laser assembly, I: laser current controller, Isolator: fiber-integrated optical isolator, FPC: fiber-integrated polarization controller, EDFA: Erbium-doped fiber amplifier, PM: power meter, ESA: electrical spectrum analyser, MZM: fiber-integrated Mach-Zehnder Modulator, AWG: arbitrary waveform generator, LiCOF: liquid core optical fiber (target), C: fiber-integrated optical circulator, PD: photo diode, heater: electrical heater plate, BD: beam dump, BPF: band-pass filter for ASE suppression.}
\label{fig:Setup_integrated}
\end{figure}

The simplified setup for the integrated SBS measurements is presented in \autoref{fig:Setup_integrated}.
A laser at wavelength $\lambda = \SI{1549}{\nano\meter}$ is split up into a local oscillator (LO) part and a pump part. 
The pump part is modulated via a Mach-Zehnder Modulator (MZM) into \SI{30}{\nano\second} pulses and subsequently amplified by an Erbium-doped fiber amplifier (EDFA).
In order to suppress the amplified spontaneous emission (ASE) a \SI{1}{\nano\meter} optical band-pass filter (BPF) is used.
The final pump power is measured using a power meter (PM). 
Via an optical circulator (C), the pump light is send into the LiCOF, where is creates back-scattered SBS employing thermal phonons as seed.
While the transmitted pump is dumped, the backscattered SBS is guided onto a high frequency photo diode via the circulator.
By mixing the SBS signal with the previously detached LO, a heterodyne measurement is performed on the electrical spectrum analyzer (ESA).
Before the MZM and before the mixing the polarization it controlled via a fiber-integrated polarization controller (FPC) in order to maximize transmission and beating amplitude, respectively.

\subsection{Details on integrated SBS measurements and simulations}
The backward integrated SBS measurements are performed at $\lambda = \SI{1549}{\nano\meter}$ using an optically heterodyne measurement technique and an electronic spectrum analyser as described in \autoref{sec:LiCOF_int_setup}.
The linewidth is extracted from seeded SBS measurements at the same wavelength with a lock-in detection for a modulated pump.
As the fiber has a core diameter of \SI{1.125}{\micro\meter}, one finds a V-number of $1.51$, indicating optical single mode operation (i.e. $V < 2.405$). 
For Numerical simulations COMSOL Multiphysics\textsuperscript{\textregistered} is employed to calculate the optoacoustic overlap between the optical fundamental mode and the first three radially symmetric longitudinal pressure acoustic modes $\mathrm{L_{01}}$, $\mathrm{L_{02}}$ and $\mathrm{L_{03}}$. Here we follow the theoretical work of Kobyakov et~al.~\cite{Kobyakov.2010.Adv.Opt.Photon.}. We assumed a constant linewidth of \SI{65}{\mega\hertz} for all modes obtained from the measurements.
To match the simulations with the measurements free parameter in the model are fitted.
The absolute frequency position is influenced by the unknown initial pressure assuming a linear relation between pressure and Brillouin frequency shift. 
The relative frequency position is matched by fitting fiber diameter and is found to be $\SI{1.125}{\micro\meter}$.
In order to fit the gain, the thermal Stokes seed as well as global gain are fitted.
Each measurement is performed once the fiber-heater system has reached thermodynamic equilibrium, which takes about five to ten minutes after changing the temperature. 

\subsection{Details on BOCDA measurements in LiCOF}

\begin{figure}[ht]
\centering
\includegraphics[width = 0.6\textwidth]{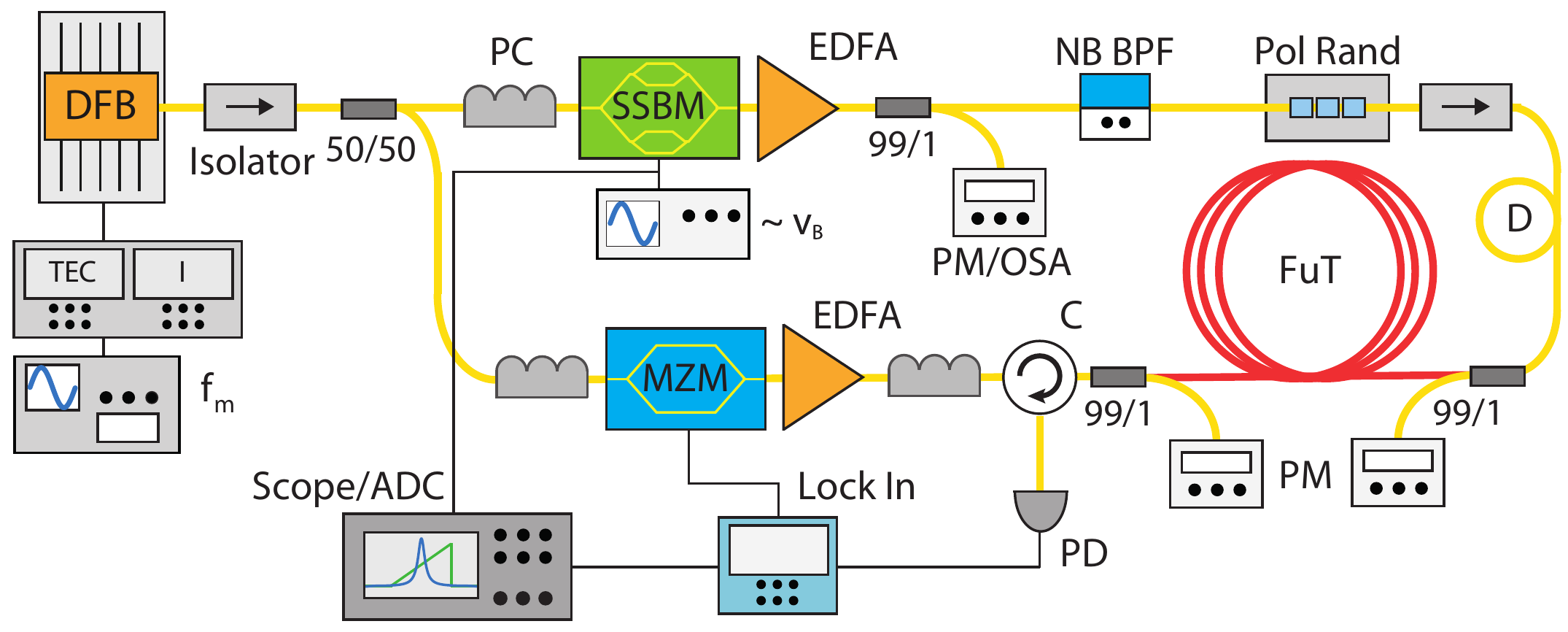}
\caption{Schematic sketch of BOCDA setup. DFB: distributed fiber Bragg laser, TEC: laser temperature controller, I: laser current controller, Isolator: fiber-integrated optical isolator, PC: fiber-integrated polarization controller, $f_m$: radio-frequency generator for laser current modulation, 50/50: fiber-integrated beamsplitter ratio 50:50, SSBM: fiber-integrated single-sideband modulator, $v_B$: radio-frequency generator for Brillouin resonance modulation, EDFA: Erbium-doped fiber amplifier, 99/1: fiber-integrated beamsplitter ratio 99/1, NB BPF: fiber-integrated narrow-band bandpass filter, Pol Rand: fiber-integrated polarization scrambler, PM: power meter, OSA: optical spectrum analyser, D: delay, MZM: fiber-integrated Mach-Zehnder Modulator, FuT: fiber under test (target), C: fiber-integrated optical circulator, PD: photo diode, Lock In: lock-in amplifier, Scope/ADC: oscilloscope or analogue-to-digital-converter.}
\label{fig:Setup_BOCDA}
\end{figure}

The complete BOCDA setup is presented in Figure \ref{fig:Setup_BOCDA}. The local acoustic response is created through a frequency modulation of the master laser source with modulation frequency $f_m = 699\,$kHz and modulation bandwidth $\Delta f = 47\,$GHz resulting in a resolution of $\Delta z = 5.9\,$cm. Due to the repetitive nature of the correlation approach, the sensing distance is limited by $d_m = {v_g}/{2f_m}$ with the group velocity $v_g$. This modulation is realized using a directly current modulated distributed fiber Bragg laser. The modulation is generated using an external RF source. After passing an isolator, the signal is split 50:50 into a pump and probe beam. The probe is passed through a single-sideband modulator, shifting it downward by the Brillouin frequency $\nu_B$. After amplification through an Erbium-doped fiber amplifier (EDFA), 1$\%$ is split off for power or spectral monitoring. The signal is filtered using a narrow-band bandpass and afterwards the polarization is scrambled with a frequency of $f_\text{rand} = 700\,$kHz. After passing a delay it is launched into the fiber under test (FuT). The seed input is monitored using a second tap-off. The pump beam is modulated with the lock-in frequency of $f_\text{lock} = 100\,$kHz. Then it is amplified using a second EDFA. After polarization adjustment the pump is launched into the FuT using a fiber-integrated optical circulator. Again the power is monitored using a 1$\%$ tap-off. The backscattered Brillouin signal is detected on a photo diode and demodulated through the lock-in amplifier. The signal is monitored at an oscilloscope with temporal reference from the single-sideband modulation frequency to retrieve the spectral information. \\

For our numerical simulations we follow the frequency domain approach of reference \cite{Yamauchi.2004.Fiber.Opt.Sens.Tech.} in combination with the non-instantaneous amplitude-to-frequency response approach of reference \cite{Song.2018.OptLett}. The discrete beat spectrum $S_{B}(z,f)$ of pump and probe is calculated and convoluted with the analytical amplitude beat spectrum $S_{A}(z,f)$ resulting from the non-instantaneous amplitude-to-frequency response. The resulting beat spectrum is finally convoluted with the Brillouin gain spectrum $g_B(f)$ to retrieve the position dependent normalized Brillouin gain $g_B(z,f)$. Due to the small resolution $\Delta z = 6.4\,$cm, the distributed Brillouin signal can be seen as a Brillouin signal the small gain limit.

\section*{\label{sec:Acknowledgments } Acknowledgments}
We acknowledge our co-workers D. Walter, F. Sedlmeir, A. Zarifi, X. Zeng, S. Becker and A. Tuniz for valuable discussions, and L. Meier,  O. Bittel, R. Gall, M. Schwab and A. Wambsganß for support with the custom built mechanics and electronics of the systems.

\section*{Funding}
We acknowledge funding from the Max Planck Society through the Independent Max Planck Research Group scheme. M.C. acknowledges funding from the Banting foundation (Canada) as well as the FRQNT through the PBEEE fellowship.

\section*{Author contributions}

B.S., M.C. and M.S. conceived the idea. A.G., A.P. and D.D performed the experiments and analyzed the data. S.J. fabricated the sample. All authors discussed the results and worked on their interpretation. A.G., A.P., M.C., C.P., M.S. and B.S. wrote the manuscript with input from all authors. M.C., C.M., M.S. and B.S. supervised the project.

\bibliography{LIB}

\end{document}


\title{Supplementary Material: Exploring extreme thermodynamics in nanoliter volumes through stimulated Brillouin-Mandelstam scattering}

\author{ Andreas Geilen $^{1,2,}$\footnote[1]{These authors contributed equally.}}

\author{Alexandra Popp$^{1,2,3,\rm{a}}$}
\author{Debayan Das$^{1,4}$} 
\author{Saher Junaid$^{5,6}$}
\author{Christopher G. Poulton$^{7}$} 
\author{Mario Chemnitz$^{6,8}$}
\author{Christoph Marquardt$^{2,1,3}$}
\author{Markus A. Schmidt$^{5,6}$}
\author{Birgit Stiller$^{1,2}$}
\email{birgit.stiller@mpl.mpg.de}

\address{
$^{1}$Max Planck Institute for the Science of Light, Staudtstr. 2, 91058 Erlangen, Germany \\
$^{2}$Department of Physics, University of Erlangen-N\"urnberg, Staudtstr. 7, 91058 Erlangen, Germany \\
$^{3}$SAOT, Graduate School in Advanced Optical Technologies, Paul-Gordan-Str. 6, 91052 Erlangen, Germany\\
$^{4}$Universit\'e Bourgogne France-Comt\'e, 25030 Besan\c{c}on, France\\
$^{5}$Leibniz Institute of Photonic Technology, Albert-Einstein-Str. 9, 07745 Jena, Germany\\
$^{6}$Otto Schott Institute of Materials Research (OSIM), Fraunhoferstr. 6, 07743 Jena, Germany \\
$^{7}$School of Mathematical and Physical Sciences, University of Technology Sydney, NSW 2007, Australia\\
$^{8}$ INRS-EMT, 1650 Boulevard Lionel-Boulet, Varennes, Québec, J3X 1S2, Canada}

\date{\today}

\maketitle

\section{Gain measurement}

\noindent In the stimulated regime, the Stokes component of Brillouin scattering follows the exponential behavior $P_\mathrm{Peak} \propto \exp(G \cdot P_\mathrm{R})$, with $G$ being the Brillouin gain and $P_\mathrm{R} = L\cdot R_\mathrm{P}$ being the reduced power. 
Here $L = \SI{2\pm0.05}{\meter}$ is the length of the capillary and $R_\mathrm{P} = 0.03$ is the duty cycle of the used pump pulses with a repetition rate of \SI{1}{\mega\hertz}.
Considering the reduced power instead of the average input power removes the dependence of the gain on experimental parameters, such as the length of the sample and the used pump scheme.
Only data points above a threshold, where the Brillouin intensity start to rise exponentially, are considered for the fitting. 
The data point around \SI{1.0}{\watt\meter} is lower as we were running into the saturation of the photo diode before introducing an optical attenuator to measure higher pump powers.
A similar behavior is observed around \SI{1.5}{\watt\meter}.

\begin{figure}[ht]
\centering
\includegraphics[width = 0.5\textwidth]{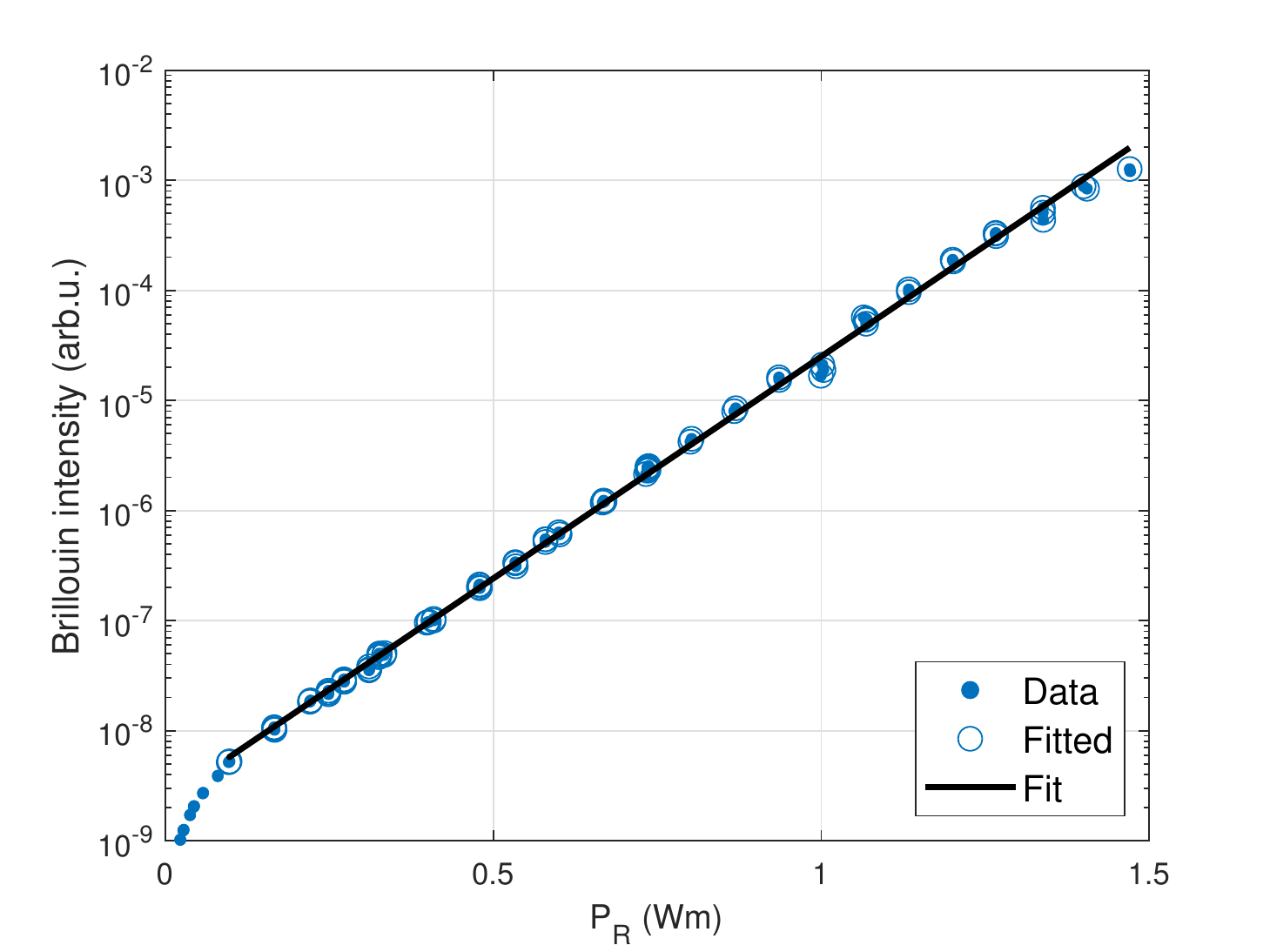}
\caption{ Measured Brillouin intensity over average pump power on logarithmic scale. Stimulated Brillouin scattering increases exponentially with the pump power. Exponential model fitted. Encircled data points are in the stimulated regime and therefore used for the fit.}
\label{fig:Gain_measurement}
\end{figure}

\section{Temperature-pressure diagram}

\noindent 

\begin{figure}[ht]
\centering
\includegraphics[width = 0.6\textwidth]{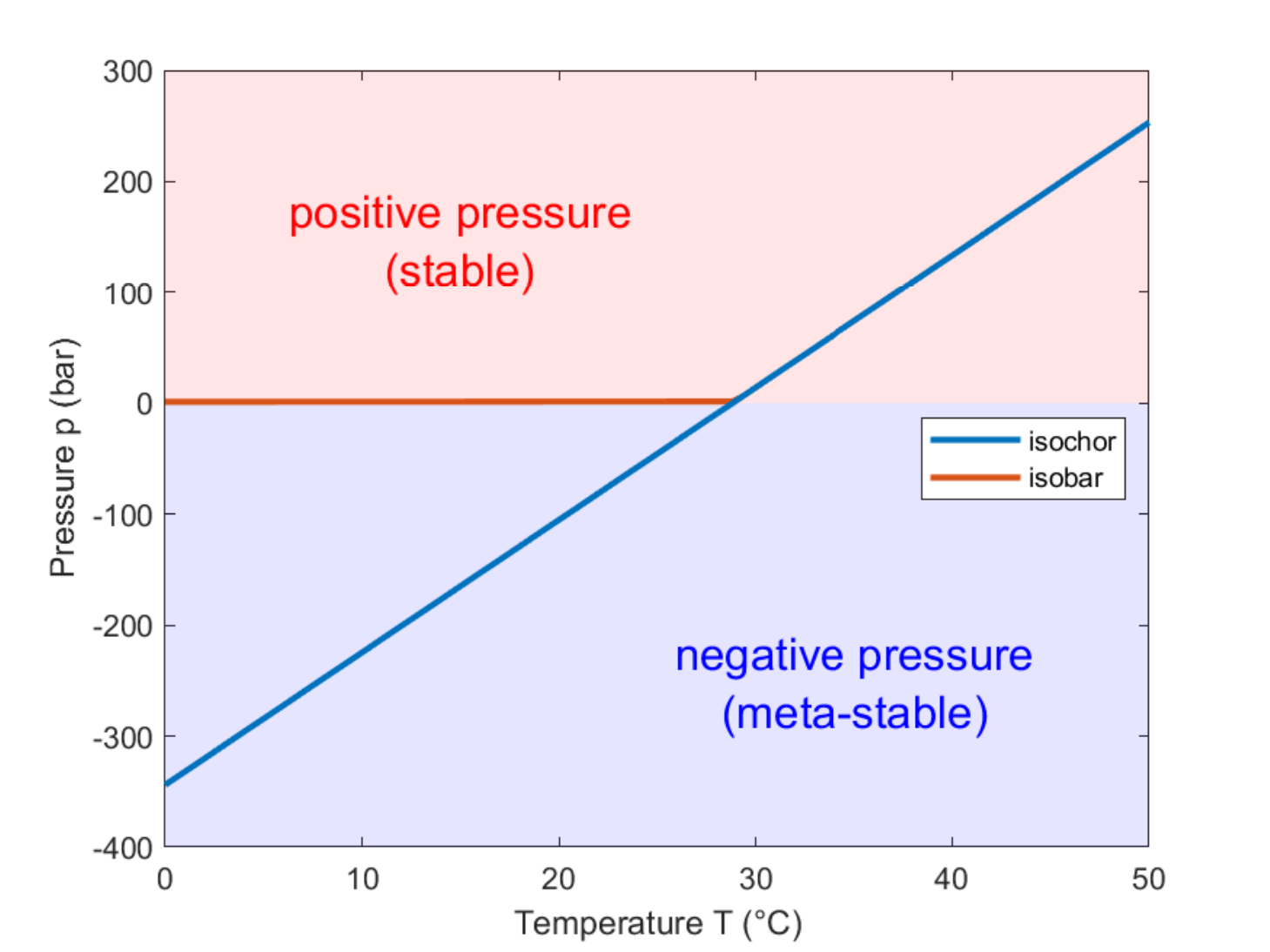}
\caption{Illustration of the isobaric and isochoric regime in the temperature-pressure diagram for a fully tempered fiber, that is \SI{99}{\percent}-filled at room temperature. In the isobaric regime, only the \ce{CS2} vapour pressure is considered, which is neglectable compared to the pressure in the isochoric state.}
\label{fig:T-p_diagram}
\end{figure}